\documentclass{ws-ijmpcs}

\begin{document}

\markboth{E. Torresi, P. Grandi, E. Costantini, G.G.C. Palumbo}
{Jets and outflows in Radio Galaxies: implications for AGN feedback}

%
\catchline{}{}{}{}{}
%


\title{JETS AND OUTFLOWS IN 
RADIO GALAXIES: IMPLICATIONS
FOR AGN FEEDBACK}

\author{ELEONORA TORRESI}

\address{INAF/IASF--Bologna\\
via Gobetti 101, 40129, Bologna,
Italy\\
torresi@iasfbo.inaf.it}

\author{PAOLA GRANDI}

\address{INAF/IASF--Bologna\\
via Gobetti 101, 40129, Bologna, Italy\\
grandi@iasfbo.inaf.it}

\author{ELISA COSTANTINI}

\address{SRON, Netherlands Institute for Space Research\\
Sorbonnelaan 2, 3584 CA Utrecht, The Netherlands\\
E.Costantini@sron.nl}

\author{GIORGIO G.C. PALUMBO}

\address{Dipartimento di Astronomia, Universit\`a di Bologna\\
via Ranzani 1, I-40127 Bologna, Italy\\
giorgio.palumbo@unibo.it}

\maketitle

\begin{history}
\received{Day Month Year}
\revised{Day Month Year}
\end{history}

\begin{abstract}
One of the main debated astrophysical problems is the role of the AGN feedback in galaxy formation. 
It is known that massive black holes have a profound effect  on the formation and evolution of galaxies, but how black holes and galaxies communicate is still an unsolved problem. For Radio Galaxies, feedback studies have mainly focused on jet/cavity systems in the most massive and X--ray luminous galaxy clusters. The recent high--resolution detection of warm absorbers in some Broad Line Radio Galaxies allow us to investigate the interplay between the nuclear engine and the surrounding medium from a different perspective.  
We report on the detection of warm absorbers in two Broad Line Radio Galaxies, 3C~382 and 3C~390.3, and discuss the physical and energetic properties of the absorbing gas.  Finally, we attempt a comparison between radio--loud and radio--quiet outflows.

\end{abstract}

\ccode{PACS numbers: 90, 95.30.Dr, 98.54.Gr}

\section{Introduction}	
In the last few years, high--resolution X--ray spectroscopy has made progress in the exploration of the circumnuclear environment of radio--loud (RL) AGNs.
While the presence of X--ray emitting and absorbing gas is well established in Seyfert galaxies, for RL sources  the investigation of the nuclear environment through this technique is very recent. Specifically, in Broad Line Radio Galaxies (BLRG), the RL counterpart of Seyfert 1s, the detection of warm absorbers (WA) \footnote{With the term ``warm absorber'' we intend ionized outflowing gas in our line--of--sight that produces narrow absorption lines in the soft X--ray spectrum.}  was expected to be more difficult because of their small number in the local Universe and  because the Doppler amplification of the jet emission could mask the absorption features. 
However, steps forward have been recently made thanks to high--resolution X--ray spectroscopy.
Here we summarize the results concerning the discovery of WAs in two BLRGs, 3C~382 and 3C~390.3.

\section{RGS spectral analysis and results}
\begin{figure}[pb]
\begin{center}
\psfig{file=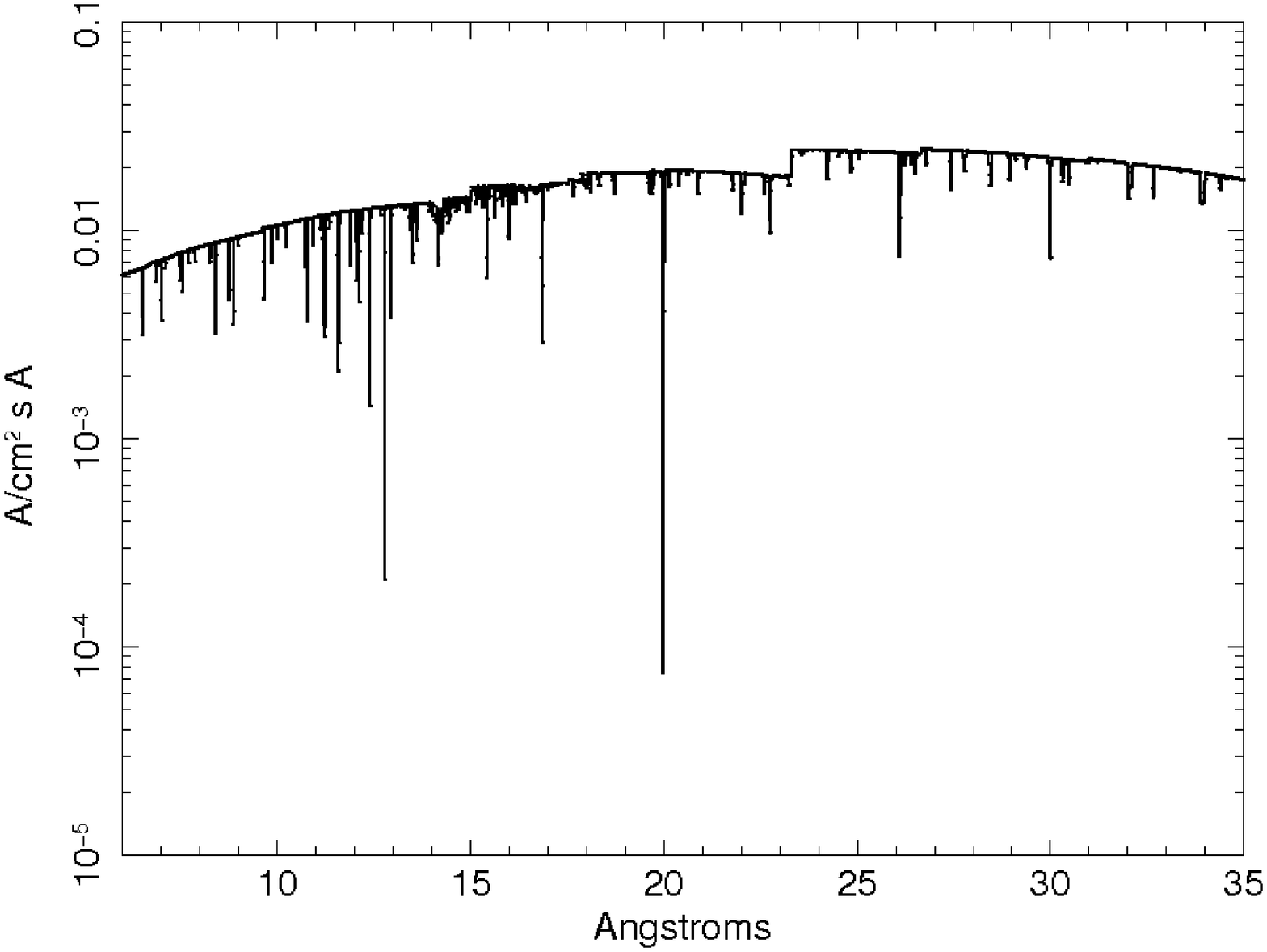,width=3.0 cm, angle=270}
\psfig{file=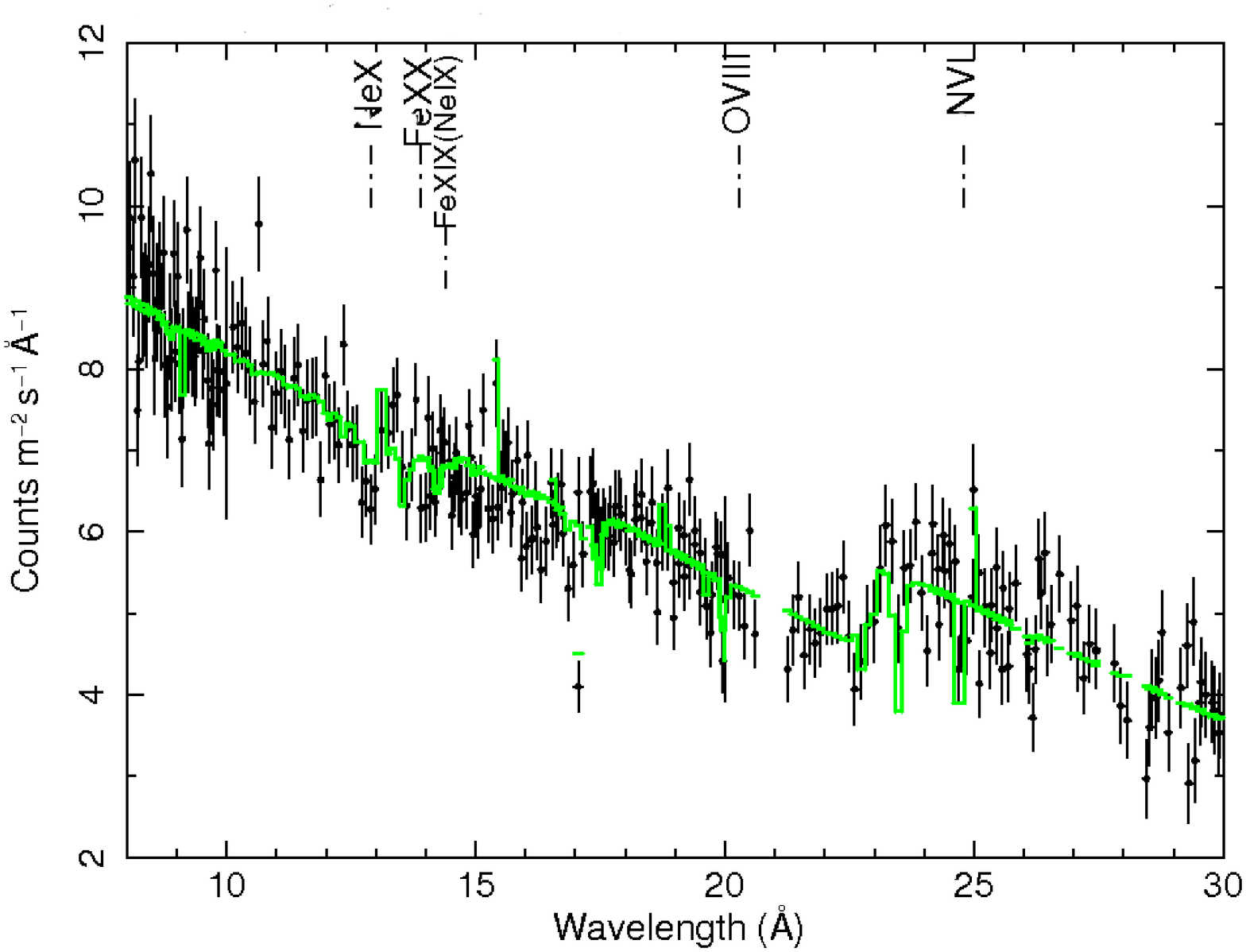,width=3.0 cm, angle=270}
\vspace*{8pt}
\caption{{\it Left panel}: XSTAR best--fit model of 3C~382. {\it Right panel}: SPEX {\it xabs} best--fit model of 3C~390.3. The most prominent absorption lines are labelled. \label{f1}}
\end{center}
\end{figure}

We analyzed all the XMM--Newton/Reflection Grating Spectrometer (RGS) data available for both 3C~382 and 3C~390.3 \cite{3c382}$^{-}$\cite{blrg}. 
Two different photoionization codes (XSTAR \cite{xstar} and {\it xabs} in SPEX \cite{spex}) have been used to model the absorption features  (Fig.~\ref{f1}) and to derive the physical parameters of the outflow, i.e. column density, ionization parameter \footnote{$\xi$=$\displaystyle\frac{L}{n_{e}R^{2}}$, \textit{L} is the 1-1000 Rydberg (Ry) source ionizing luminosity (corresponding to 13.6 eV--13.6 keV), \textit{n$_{e}$} is the electron density of the gas and \textit{R} is the distance of the gas from the central source.} and outflow velocity.
The absorbing gas is highly ionized in both sources with log$\xi >$2~erg~cm~s$^{-1}$, column densities varying in the range N$_{H}$$=$10$^{20-22}$~cm$^{-2}$ and  outflow velocities v$_{\rm out}$$\sim$10$^{2-3}$~km~s$^{-1}$. These slow velocities constrain the location of such gas between the torus and the NLR, favoring the torus wind scenario \cite{kk}$^{-}$\cite{b05}. \\

\newpage
\noindent
Table~1 summarizes the WA energetics.

\begin{table}[ph]
\tbl{WA energetics of 3C~390.3 and 3C~382. The mass outflow rate, the kinetic energy related to the outflow, the accretion luminosity, the mass accretion rate and the jet kinetic power are listed.}
{\begin{tabular}{@{}lccccc@{}} \toprule
& log$\dot{\rm M}_{\rm out}$ & log$\dot{\rm E}_{\rm out}$ & logL$_{\rm acc}$ & $\dot{\rm M}_{\rm acc}$ & logP$_{\rm jet}$   \\
 &  (M$_{\odot}$~yr$^{-1}$) & (erg~s$^{-1}$) & (erg~s$^{-1}$) &(M$_{\odot}$~yr$^{-1}$)  & (erg~s$^{-1}$)   \\ \colrule
3C~390.3& 1.40\hphantom{00} & \hphantom{0}41.66 & \hphantom{0}45.65 & 0.77& 45.12 \\
3C~382& 1.41\hphantom{00} & \hphantom{0}41.91 & \hphantom{0}45.84 & 1.2& 44.81\\ \botrule
\end{tabular} \label{ta1}}
\end{table}

\noindent
The mass outflow rate ($\dot{\rm M}_{\rm out}$) estimates the mass carried out of the AGN through the wind \footnote{$\dot{M}_{out}$$\sim$$\frac{1.23  m_{p} L_{ion} v_{out} C_{v}\Omega}{\xi} $, the solid angle is set to $\Omega$=2.1, while the volume filling factor (C$_{\rm v}$) is kept equal to 1.}, while the WA kinetic energy ($\dot{\rm E}_{\rm out}$) is the power released in the circumnuclear environment through the outflow \footnote{$\dot{E}_{out}$=$\frac{\dot{M}_{out}v_{out}^{2}}{2}$.}. P$_{\rm jet}$ \footnote{$P_{jet}=3\times10^{45}f^{3/2}L^{6/7}_{151} ~{\rm erg~s}^{-1} $, where L$_{151}$ is the observed radio luminosity at 151 MHz and $<$f$>$=15 accounts for systematic underestimates of the true jet power  \cite{hardcastle} .} is the jet kinetic power calculated according to the formula of Ref.~\refcite{shankar}.
Note that considering C$_{\rm v}$=1, the mass outflow rates are implausibly higher than the mass accretion rates, implying a clumpy configuration of the gas (C$_{v}<$1). Indeed, assuming for our sources that the same amount of matter is accreted and ejected in the form of wind, we can deduce a volume filling factor as small as $\sim$0.01.
Moreover,  from Table~1 it is evident that the kinetic luminosity related to these slow outflows is a negligible fraction ($<<$1$\%$) of both bolometric luminosity and jet kinetic power.

\section{Comparison with type 1 Radio--Quiet AGNs}
\begin{figure}[pbh]
\begin{center}
\psfig{file=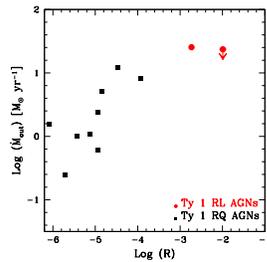,width=3.7 cm, angle=0}
\vspace*{8pt}
\caption{Mass outflow rate plotted against the radio--loudness parameter (R). {\it Red circles}: BLRGs considered in this work; {\it black squares}: type 1 RQ AGNs belonging to our reference sample. \label{f2}}
\end{center}
\end{figure}

Aware of the scarcity of RL sources with WAs, we attempt a comparison between their X--ray properties with a sample of type 1 RQ AGNs (Seyfert 1s, NLS1s, QSOs) having a good modeling of the absorption features  \cite{b05}. 
Again, fixing C$_{\rm v}$=1 the mass outflow rates have implausibly large values in both RL and RQ objects, independently of their radio power. Also the kinetic luminosity related to the slow outflows is negligible with respect to the accretion luminosity.  
Finally, in order to investigate the role of the relativistic jet we explore a possible correlation between the mass outflow rate and the radio--loudness parameter (R) \footnote{$R=Log[ \frac{\nu L_{\nu(5GHz)}}{L_{(2-10keV)}}]$ as proposed by Ref.~\refcite{rl}.}. 
Looking at Fig.~\ref{f2} a possible positive correlation between $\dot{\rm M}_{\rm out}$ and R can be observed. This trend could suggest a different distribution of the gas in RL and RQ sources, tending to preferentially clump when the system is less perturbed by the jet. Alternatively, if the geometry of the gas is similar in both classes, such trend could indicate that larger amount of mass escapes from the central engine when a powerful jet is present.

\section{Summary}
We report on the detection of WAs in two BLRGs, 3C~382 and 3C~390.3, and discuss the physical and energetic properties of the absorbing gas:
{\bf (i)} the outflows are highly ionized (log$\xi >$2~erg~cm~s$^{-1}$) and slow, with velocities ranging between 10$^{2}$--10$^{3}$~km~s$^{-1}$;
{\bf (ii)} the mass outflow rates are higher than the mass accretion rates if a volume filling factor (C$_{\rm v}$) equal to 1 is assumed. Therefore a gas clumpy configuration (C$_{\rm v} <$1) is expected;
{\bf (iii)} the kinetic luminosity associated to these slow outflows is always lower than the accretion luminosity and the jet kinetic power;
{\bf (iv)} although RL and RQ WA physical properties appear very similar, at least at zeroth order, the mass outflow rate and the radio--loudness parameter (R) seem to be correlated. This correlation could indicate a different gas distribution or alternatively, if the gas distribution is the same, powerful jets could favor the escape of more massive winds. 

\section*{Acknowledgments}
ET acknowledges the support of the Italian Space Agency (contract ASI/INAF I/009/10/0 and ASI/GLAST I/017/07/0).


\end{document}